# Geometry effects on zonal flow dynamics and turbulent transport in optimized stellarators

Haotian Chen[1,2], Xishuo Wei[2], Hongxuan Zhu[3], and Zhihong Lin[2,*]

[1]Fusion Simulation Center, Peking University, Beijing 100871, China
[2]Department of Physics and Astronomy, University of California, Irvine, CA 92697, USA
[3]Department of Astrophysical Sciences, Princeton University, Princeton, NJ 08544, USA

Global gyrokinetic simulations find a strong suppression of ion temperature gradient (ITG) turbulence by zonal flows in stellarators optimized for neoclassical transport. The reduction of the ITG transport by the zonal flows in quasi-helicalsymmetric (QH) and quasi-isodynamic (QI) stellarators are much larger than a quasi-axisymmetric (QA) stellarator or a tokamak, thanks to higher linear residual levels and lower nonlinear frequencies of the zonal flows in the QH and QI. The transport level and energy confinement time in the QH and QI are similar to the tokamak with the same size and temperature gradient, despite the much larger linear growth rates in the stellarators.

Stellarators with 3-dimensional (3D) magnetic configurations [1] are an attractive fusion reactor concept thanks to their steady state operation and reduced risk of disruptions since no plasma current drive is needed [2]. Advances in stellarator design have drastically reduced the collisional (so-called neoclassical) transport in the optimized stellarators such as quasi-isodynamic (QI) stellarators [3-5] and quasi-symmetric stellarators [6]. With the neoclassical transport reduced, turbulent transport driven by driftwave instability can dominate the particle and heat transport in the optimized stellarators such as the W7-X experiments [7]. To be a competitive reactor candidate, the optimized stellarators need to demonstrate a turbulent transport level similar to or lower than an axisymmetric tokamak. While turbulent transport in the optimized stellarators has recently been studied using gyrokinetic simulations [8-12] and some comparisons of instability and transport between different geometries have been made [12-14], comprehensive studies of turbulent transport across all optimized stellarators and tokamaks have yet to be performed.

In this work, global gyrokinetic simulations find that turbulent transport driven by the electrostatic ion temperature gradient (ITG) instability in the model equilibria of the QI [10] and quasi-helical symmetric (QH) [15] stellarators is at a level similar to that in an ITER tokamak scenario [16], despite a much larger linear growth rate in the stellarators. The underlying physics is found to be the reduction of turbulent transport by zonal flows, which have much higher linear residual levels and lower nonlinear frequencies in the QI and QH stellarators than the tokamak or the quasi-axisymmetric (QA) [15, 17] stellarator. These results demonstrate the potential of the optimized stellarators as a promising reactor candidate and suggest a new research direction by optimizing zonal flow dynamics for the turbulence self-regulation to improve plasma confinement in the design of stellarator reactors.

*Global gyrokinetic simulations--* Four model equilibria of optimized stellarators have been simulated in the current study: a QA and a QH recently optimized for neoclassical transport [15], a QI based on a W7-X model equilibrium [10], and a compact QA design NCSX optimized for ballooning stabilization [17]. The equilibria of these toroidal geometries have been calculated by the equilibrium code VMEC [18] and cast in Boozer coordinates $(\psi, \theta, \zeta)$, where $\psi$ is poloidal flux, $\theta$ is poloidal angle, and $\zeta$ is toroidal angle [19]. For comparison, a model tokamak equilibrium of an ITER steady state scenario [16] has also been simulated. For fair comparisons of transport levels, all devices have the same minor radius and temperature gradient. The minor radial coordinate $r$ is defined as $r = V/2\pi^2 R$, where $R$ is toroidally averaged major radius, $V(\psi)$ is the volume inside a flux surface $\psi$. The minor radius is defined as $a = r(\psi = \psi_X)$ at the separatrix $\psi_X$. We assume that ions and electrons have the same temperature $T_i = T_e = T(r)$ with a Maxwellian distribution and a uniform density $n$. The temperature gradient is represented by an inversed scale length $L_T^{-1} = -dlnT/dr$. The temperature gradient is uniform for the radial domain $r/a$=[0.3,0,8] and decreases within a width of about $0.1a$ to zero outside this domain. Key parameters of all devices are listed in Table 1 including field period, aspect ratio $R/a$, average safety factor $\langle q \rangle$, elongation, and Rosenbluth-Hinton (*RH*) zonal flow residual [20]. Here, elongation $= max\{a_{max}/a_{min}\}_\zeta$, $a_{max}$ and $a_{min}$ are, respectively, the maximal and minimal width of the boundary shape at a toroidal angle and $max\{...\}_\zeta$ means taking the maximal value across all toroidal angles. All devices have the same size of $a = 124\rho_s$ with $\rho_s = C_s/\Omega_i$, ion cyclotron frequency $\Omega_i$, ion sound speed $C_s = \sqrt{T_0/m_i}$, ion mass $m_i$, and on-axis temperature $T_0$. The *RH* residuals are calculated in simulations of only zonal flows with $k_r\rho_s = 0.5 - 1$, which are dominant zonal flow wavevectors in the ITG turbulence [21].

TABLE 1. Key parameters of all simulated devices.

|  | QA | NCSX | QH | QI | tokamak |
|---|---|---|---|---|---|
| Period | 2 | 3 | 4 | 5 | N/A |
| $R/a$ | 6.2 | 4.5 | 6.8 | 12.4 | 2.4 |
| $\langle q \rangle$ | 2.39 | 1.80 | 0.94 | 1.12 | 2.10 |
| Elongation | 4.62 | 2.71 | 3.52 | 2.74 | 1.80 |
| *RH* | 0.056 | 0.11 | 0.47 | 0.21 | 0.14 |

The simulations of these devices have been performed using the global gyrokinetic code GTC [22] with an efficient field-aligned mesh in the real space [23] and effective utilization of the most powerful supercomputers [24]. Global simulations are needed for the 3D stellarator geometry to incorporate linear and nonlinear couplings of toroidal harmonics, turbulence spreading, and full flux-surface-averaged zonal flow dynamics [12]. GTC has recently been applied to study ITG turbulence [11] and collisionless damping of zonal flows [21] in the LHD and W7-X stellarators, neoclassical and turbulent transport with self-consistent ambipolar electric fields in the W7-X [25], and ITG and trapped electron mode using kinetic electrons in the LHD [26, 27]. In the current simulations, ions are governed by the nonlinear gyrokinetic equation [28] and electron responses are assumed to be adiabatic. Perturbed electrostatic potential is calculated using the gyrokinetic Poisson equation [29].

Taking advantage of the stellarator field periods, only one period of the torus is simulated for the stellarators, while the full torus is simulated for the tokamak. Convergence studies on spatial grid size, time step, number of particles per cell, and radial boundary treatment have been successfully conducted. Based on these convergence studies, the current simulations use (120-240, 3600-4800, 27) grid points in, respectively, radial, poloidal, and parallel directions. The time step size is $\Delta t = 0.016\text{-}0.105 L_T/C_s$ and 200-400 particles per cell are used to minimize the noise in nonlinear simulations.

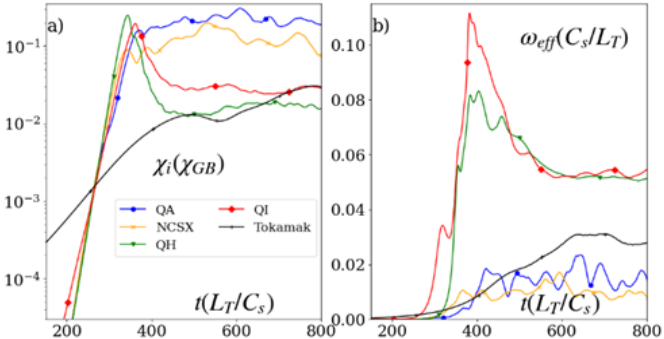

FIG. 1. Time history of heat conductivity $\chi_i$ (panel a) and zonal flow effective shearing rate $\omega_{eff}$ (panel b).

*Comparisons of turbulent transport levels—* We first performed simulations with a realistic temperature gradient of $a/L_T = 1.4$. The unstable ITG modes grow exponentially in the linear phases and then saturate to a quasi-steady state in the nonlinear phase, as shown in FIG. 1 for the time history of ion heat conductivities and zonal flow shearing rates. Here the heat conductivity is defined as $\chi_i \equiv QL_T/nT$ and is normalized to gyroBohm unit $\chi_{GB} \equiv \rho_s^2 C_s/L_T$. The heat flux density is measured by $Q = \langle \int (E - 1.5T) v_{dr} \delta f d\mathbf{v} \rangle$. Here $E$ is particle kinetic energy, $v_{dr}$ is radial ExB drift due to perturbed electric fields, $\int d\mathbf{v}$ is velocity space integral with perturbed distribution function $\delta f$, and $\langle \ \rangle$ represents average over the flux-surface and a radial width of full-width-at-half-maximum of the linear mode structure. The zonal flow shearing rate [30] is defined as $\omega_E \equiv (\partial^2 \phi_{zf}/\partial \psi^2)(\Delta r/\Delta \theta) RB_\theta/q$, where $\phi_{zf}$ is the zonal electrostatic potential $\phi$, $\Delta r$ and $r\Delta \theta$ are, respectively, radial and poloidal correlation length (which are assumed to be same when calculating $\omega_E$), and $B_\theta$ is the poloidal magnetic field. The effective shearing rate is defined as $\omega_{eff} = \omega_E[(1 + 3F)^2 + 4F^3]^{1/4}/[(1 + F)\sqrt{1 + 4F}]$ with $F \equiv \omega_{zf}^2/\Delta \omega_T^2$. Here $\omega_{zf}$ is the frequency of the zonal flow and $\Delta \omega_T$ is the decorrelation rate of the ambient turbulence [31].

The most striking result shown in FIG.1 (a) is that the heat conductivities $\chi_i$ of the QH and QI are similar to the tokamak in the quasi-steady state, despite the much smaller linear growth rate $\gamma$ in the tokamak due to its smaller $R/L_T$ resulting in a stronger Landau damping. On the other hand, the $\chi_i$ of the QA (and NCSX) in the quasi-steady state are much higher than the QH and QI even though the linear growth rates and initial saturation levels are very similar in all stellarators. At the ITG nonlinear saturation, there are large bursts of effective shearing rates $\omega_{eff}$ in the QH and QI, but the $\omega_{eff}$ is small at the saturation in the QA and tokamak as shown in FIG. 1 (b). Consequently, the $\chi_i$ of the QH and QI drops quickly over an order of magnitude, but there is a much smaller drop of $\chi_i$ in the QA and tokamak. These differences in transport and zonal flows are consistent and indicate different zonal flow dynamics in different magnetic configurations. In the quasi-steady state turbulence, the $\omega_{Eff}$ of the QH and QI are much larger than the QA and tokamak.

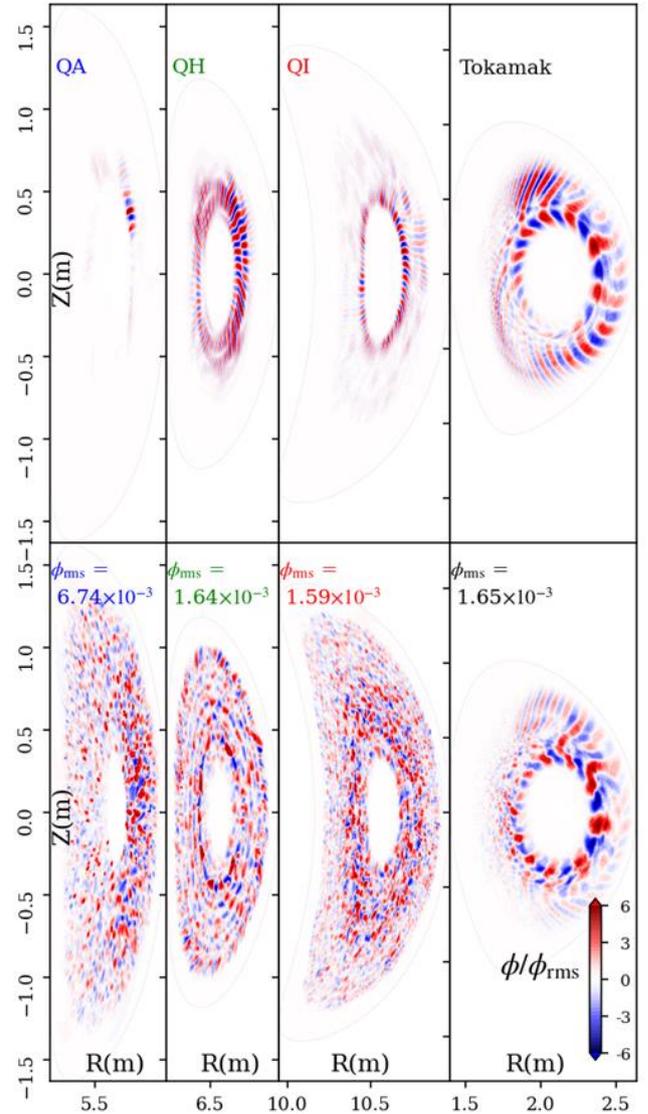

FIG. 2. Non-zonal potentials $e\phi/T$ on the poloidal plane at $\zeta = 0$ in linear (upper row, $t=250 L_T/C_s$) and nonlinear (lower row, $t=630 L_T/C_s$) phases. Nonlinear fluctuation amplitudes $\phi_{rms}$ are listed.

The shearing effects by the zonal flows are clear in FIG.2 showing perpendicular mode structures on the poloidal plane at $\zeta=0$ with stellarator up-down symmetry. The non-zonal components of perturbed electrostatic potentials $\phi$ exhibit elongated radial streamers in the linear phase (upper panels) for all devices. The mode structures on the flux-surface (not shown) are uniform across different magnetic fieldlines in the tokamak, but localized to some magnetic fieldlines in all stellarators due to the 3D equilibrium effects [11]. In the nonlinear phase (lower panels), the turbulence eddies are mostly isotropic as zonal flows break apart the linear radial streamers [22]. Meanwhile, nonlinear turbulence spreading significantly broadens the envelopes of the fluctuation intensity in both radial and poloidal directions. On the other hand, the long parallel wavelength (not shown) largely remains unchanged from the linear to nonlinear phases. In the quasi-steady state, the fluctuation amplitudes $\phi_{rms}$ in the QH, QI, and tokamak are all at similar levels, but are at much higher levels in the QA (and NCSX).

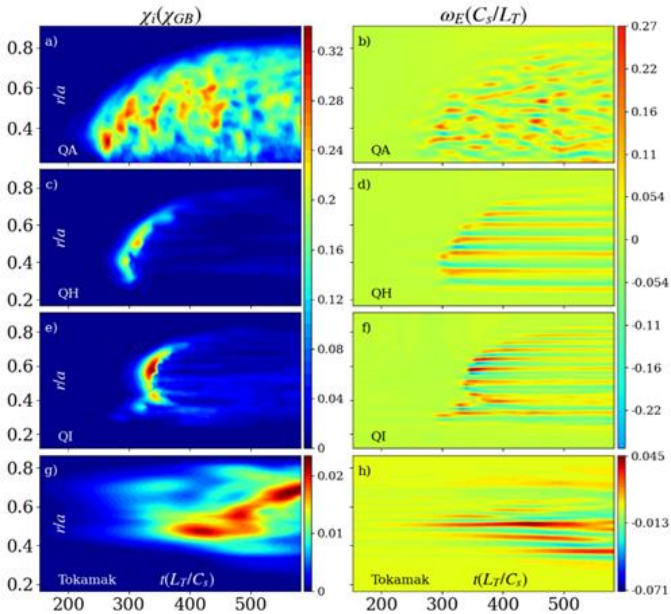

FIG.3. Time evolutions of the radial profiles of ion heat conductivities $\chi_i$ (left column) and zonal flow shearing rates $\omega_E$ (right column).

*Dynamics of zonal flows—* To understand the different shearing effects by zonal flows, we plot in FIG.3 the time evolutions of the radial profiles of ion heat conductivities $\chi_i$ (left column) and shearing rates $\omega_E$ (right column). Zonal flows are generated during the nonlinear ITG saturation, which is accompanied by a strong turbulence spreading in both radial and poloidal directions. The generated zonal flows are quickly damped by the collisionless magnetic pumping effects [20], resulting in a lower residual level. This quasi-static zonal flow residual then saturates the ITG instability and suppresses the turbulent transport in the quasi-steady state. Consistent with the GTC nonlinear simulations, linear GTC simulations (shown in Table 1) and gyrokinetic theory find higher residual levels thanks to the smaller zonal flow dielectric constants because of the smaller effective safety factor in the QH [32] and the smaller radial orbit width in the QI [33] than the QA and tokamak. The larger zonal flow residuals in the QH and QI are found to strongly correlate with the larger suppression of the turbulent transport [34] than that in the QA and tokamak as shown in FIG. 4 (d).

Another important feature in FIG.3 is that the nonlinear frequency of zonal flows in the QA (and NCSX) is much higher than the QH and QI. The radial structures of the zonal flows in the QH and QI are much more coherent and static than the QA, indicating a stronger nonlinear instability of zonal flows [34] in the QA. The nonlinear frequency of the zonal flows increases in other simulations using a larger $a/L_T$. Both the higher residuals and the smaller nonlinear frequencies of zonal flows in the QH and QI lead to a much more significant reduction of the turbulent transport than the QA and tokamak in the quasi-steady state. The geometry dependence of the zonal flow linear residual and nonlinear instability opens a new direction for stellarator optimization.

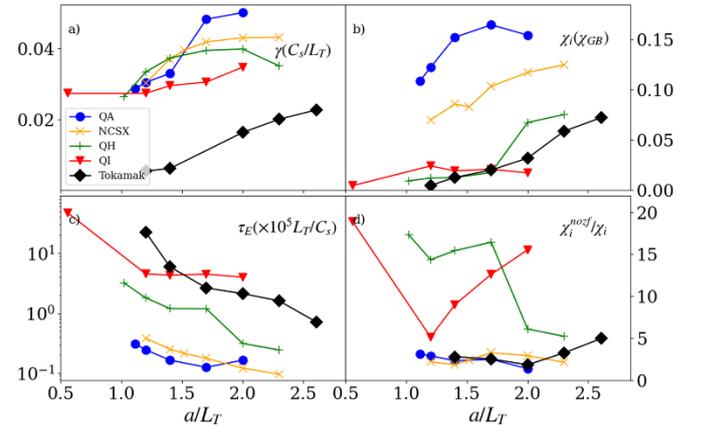

FIG. 4. Dependence of linear growth rates $\gamma$ (panel a), heat conductivity $\chi_i$ (panel b), confinement times $\tau_E$ (panel c), and reduction by zonal flows (panel d) on temperature gradients.

*Transport scaling—* We study the dependence of the transport levels and confinement times on the ITG instability drive by varying the temperature gradient while keeping all other parameters unchanged. As shown in FIG. 4 (a), the linear ITG growth rates $\gamma$ of the tokamak are much smaller than all the stellarators while real frequencies (not shown) are similar between the tokamak and stellarators. However, the ion heat conductivities $\chi_i$ of the tokamak are comparable to the QH and QI (panel b). On the other hand, the QA (and NCSX) heat conductivities are much larger than the QH, QI, and tokamak. All linear growth rates and heat conductivities in the gyroBohm units mostly increase with the temperature gradients. Consistently, all energy confinement times decrease gradually with the temperature gradients. The QH and QI energy confinement times are comparable to the tokamak, while the QA energy confinement times are much shorter (panel c). Here the energy confinement time is defined as $\tau_E = W/QS$, where $W$ is the kinetic energy confined inside the flux surface with the maximal turbulence intensity and $S$ is the area of that flux surface.

The reductions of heat conductivity $\chi_i^{nozf}/\chi_i$ by zonal flows are shown in FIG. 4 (d). Here the heat conductivities $\chi_i^{nozf}$ are measured in simulations where zonal flows are artificially suppressed. The reductions in the QH and QI are much bigger than the QA and tokamak, consistent with the lower zonal flow nonlinear frequencies and higher zonal flow residuals $RH$ in the QH and QI. The strong dependence of the transport reduction on the zonal flow residual and nonlinear frequencies indicates the importance of the ITG turbulence self-regulation by the zonal flows [22, 34].

Despite the large reduction of the ITG transport by the zonal flows, the current electrostatic simulations with adiabatic electrons produce the ion heat conductivities within the order of magnitude ranges of experimental values. For example, the ion heat conductivity for the QI case in FIG. 4(b) for the case of $a/L_T = 1.4$ corresponds to $\chi_i = 0.19 m^2 s^{-1}$ using parameters of a W7-X experiment [7], which has a measured ion heat conductivity $\chi_i = 0.25 m^2 s^{-1}$.

*Conclusion and discussions—* The current global gyrokinetic simulations find that the reduction of the ITG transport by zonal flows in the QH and QI stellarators are much larger than the QA stellarator or a tokamak, thanks to higher linear residual levels and lower nonlinear frequencies of the zonal flows in the QH and QI. The resulting transport level and energy confinement time in the QI and QH are similar to that in the tokamak with the same size and temperature gradient, despite the much larger linear growth rates in the stellarators. These findings open a new opportunity for further improving plasma confinement by maximizing the linear residual and nonlinear stability of zonal flows in the design of optimized stellarator reactors. These simulations suggest that optimized stellarators can achieve turbulent transport lower than tokamaks after further optimizing zonal flow dynamics.

Future work will study effects of kinetic electrons and electromagnetic turbulence on the confinement properties of optimized stellarators. Recently designed stellarators optimizing linear ITG drive [5, 35, 36] will be studied when equilibrium data becomes available. We will also explore the new possibility of zonal flow optimization in the design of stellarators to reduce both turbulent transport of thermal plasmas and energetic particle transport by Alfven eigenmodes [37].

This work was supported by the U.S. DOE grant DE-FG02-07ER54916, SciDAC HiFiStell, and INCITE program. Simulations used computing resources at ORNL (DOE Contract DE-AC05-00OR22725) and NERSC (DOE Contract DE-AC02-05CH11231).

*E-mail: zhihongl@uci.edu